\documentclass[aps,pra,twocolumn,showpacs,groupedaddress,amsmath,amssymb]{revtex4}

\usepackage{graphicx}
\usepackage{color}
\usepackage{comment}
\usepackage{bm}
\usepackage{latexsym}
\newcommand{\tp}[1]{\textcolor{black}{#1}}

\begin{document}

\title{Driving-induced metamorphosis of transport in arrays of coupled resonators
%Effective static Hamiltonians for driven coupled resonators
}
\author{Huanan Li, Tsampikos Kottos}
\affiliation{Department of Physics, Wesleyan University, Middletown, CT-06459, USA}
\author{Boris Shapiro}
\affiliation{Technion - Israel Institute of Technology, Technion City, Haifa 32000, Israel}
\date{\today}

\begin{abstract}
We  propose a new driving scheme, when different parts of a system are driven with different, generally incommensurate, 
frequencies. Such driving provides a flexible handle to control various properties of the system and to obtain new types of 
effective (static) Hamiltonians with arbitrary static on-site potential, be it deterministic or random. This allows us to obtain 
reconfigurable changes in transport, from ballistic to localized (including sub- and super-diffusion), depending on the driving 
protocol. The versatile reconfigurability extends also to scattering from (locally) driven extended targets. We demonstrate 
our scheme using, an analytically solvable example, of an one-dimensional tight-binding chain with appropriately driven 
couplings between nearby sites.
\end{abstract}
\maketitle

\section{Introduction}

Subjecting a quantum particle to a time dependent potential, i.e. ``driving it in time", can have a profound effect on the 
particle's dynamics. For instance, free propagation of a particle (by tunnelling from one site to the next) in a periodic tight binding lattice, 
can be inhibited by applying an ac electric field uniform in space \cite{Dun} (see \cite{Kivshar, Eckardt} for recent reviews). The effect of 
the ac field is to replace the ``bare" inter-site coupling constant $v$, before driving, by a new, effective coupling constant  $v_{eff}$. The
 latter is smaller than $v$ and, for the appropriate values of the parameters can become zero, thus, completely suppressing tunneling 
between the nearest neighbor sites. Ref. \cite{Dun} provides an early example of an effective Hamiltonian, due to driving, with properties 
different from those of the original, non-driven Hamiltonian. Recently, the same idea of driving-induced renormalized coupling has been 
utilized in the frame of non-Hermitian systems in order to manage the spontaneous ${\cal PT}$-symmetry breaking for parity-time symmetric 
systems (see \cite{CLEK17} and references therein).

Effective Hamiltonians that control the long time dynamics of a driven system, while themselves being time independent, are of great 
interest in diverse fields such as  condensed matter \cite{oka,kita,lind},  optics \cite{Kivshar, topo, topo1,topo2,topo3, gauge,gauge1,gauge2}, 
acoustics \cite{acous, acous1} and cold atomsphysics \cite{Eckardt}. For periodic in time driving the corresponding effective Hamiltonian 
is the Floquet Hamiltonian which defines the system evolution during one driving cycle. The art of obtaining new and interesting Hamiltonians 
with intriguing properties (electronic \cite{oka,kita,lind} or photonic \cite{topo, topo1, topo2, topo3} topological insulators, artificial gauge 
fields \cite{gauge, gauge1, gauge2}, time crystals \cite{time}) is sometimes called ``Floquet engineering". In addition to the cited examples, 
let us mention the effect of Floquet driving on transport in disordered one-dimensional chains \cite{hatami, Hanan, Hanan1}. In particular, 
in Ref. \cite{Hanan, Hanan1} it was demonstrated that engineering the driving spatially can largely enhance transport but it can also lead 
to a ``generalized dynamic localization", when certain parts of the chain get decoupled from the rest of it.

In this Letter, we consider a few examples of driven systems that can be {\it analytically} solved. We show that by appropriately driving in 
time the couplings between the sites of a tight- binding chain, one can create an arbitrary effective static potential, e.g., a disordered potential 
which will localize an excitation (particle or wave) on the chain. Or, on the contrary, one can ``undo'' any static disordered potential and, thus, 
let the previously localized excitation freely propagate along the chain. In fact, we shall show that our scheme allows us to design other, 
more exotic type of evolution, resulting in super-diffusion, diffusion or sub-diffusion, by an appropriate tailoring of the driving frequencies of 
the coupling constants. The new element of our driving scheme is that we allow for different (arbitrary) driving frequencies in different parts 
of the system so that, in general, we are not dealing here with a Floquet problem. Furthermore, we also extend our treatment to scattering 
problems and show that by driving the system (locally) in time, one can control scattering of the waves incident on the system.

%------------------------
\section{Model and Driving Schemes}

In our approach, we take inspiration from the exactly soluble Rabi problem which, 
in convenient for our purpose notations, amounts to
\begin{align}
\imath\frac{\mathrm{d}}{\mathrm{dt}}\begin{pmatrix}\psi_{1}\\
\psi_{2}
\end{pmatrix}= \tilde{H}\left(t\right)\begin{pmatrix}\psi_{1}\\
\psi_{2}
\end{pmatrix};
\tilde{H}\left(t\right)\equiv
& \begin{pmatrix}\omega_{1} & -ve^{\imath\omega t}\\
-ve^{-\imath\omega t} & \omega_{2}
\end{pmatrix}\label{eq: Rabi}
\end{align}
In the original Rabi problem, the $e^{\imath\omega t}$ factor comes from the circularly polarized magnetic field while in our setup it corresponds 
to driving of the coupling between two sites. In the context of optics or microwave physics the two sites can represent two coupled waveguides 
\cite{Segev} or two optical resonators (CROW), with eigenfrequencies $\omega_{1}$ and $\omega_{2}$, evanescently coupled with a coupling 
constant modulated in time \cite{Yariv}. In acoustics one can consider the scenario of two coupled Helmholtz resonators (or two cantilevers
or forks) which are coupled via air domains which experience time-dependent pressure variations.

The time-dependent gauge transformation $\psi_{1}\left(t\right)=\chi_{1}\left(t\right)e^{\imath\omega t/2}$ and $\psi_{2}\left(t\right)=\chi_{2}
\left(t\right)e^{-\imath\omega t/2}$ transforms the time-dependent Hamiltonian $\tilde{H}\left(t\right)$ into a time-independent one:
\begin{align}
\imath\frac{\mathrm{d}}{\mathrm{dt}}\begin{pmatrix}\chi_{1}\\
\chi_{2}
\end{pmatrix}= H
\begin{pmatrix}\chi_{1}\\
\chi_{2}
\end{pmatrix}; H\equiv& \left(\begin{array}{cc}
\omega_{1}+\frac{\omega}{2} & -v\\
-v & \omega_{2}-\frac{\omega}{2}
\end{array}\right)
\end{align}
Thus, driving of the type indicated in Eq.~(\ref{eq: Rabi}) (the ``Rabi driving'') can be eliminated by a gauge transformation, resulting in a 
time-independent Hamiltonian $H$ with renormalized site energies.

%---------------------------------------------------------------------------------------------------------------------------------
\begin{figure}
\begin{center}
\includegraphics[width=1\columnwidth,keepaspectratio,clip]{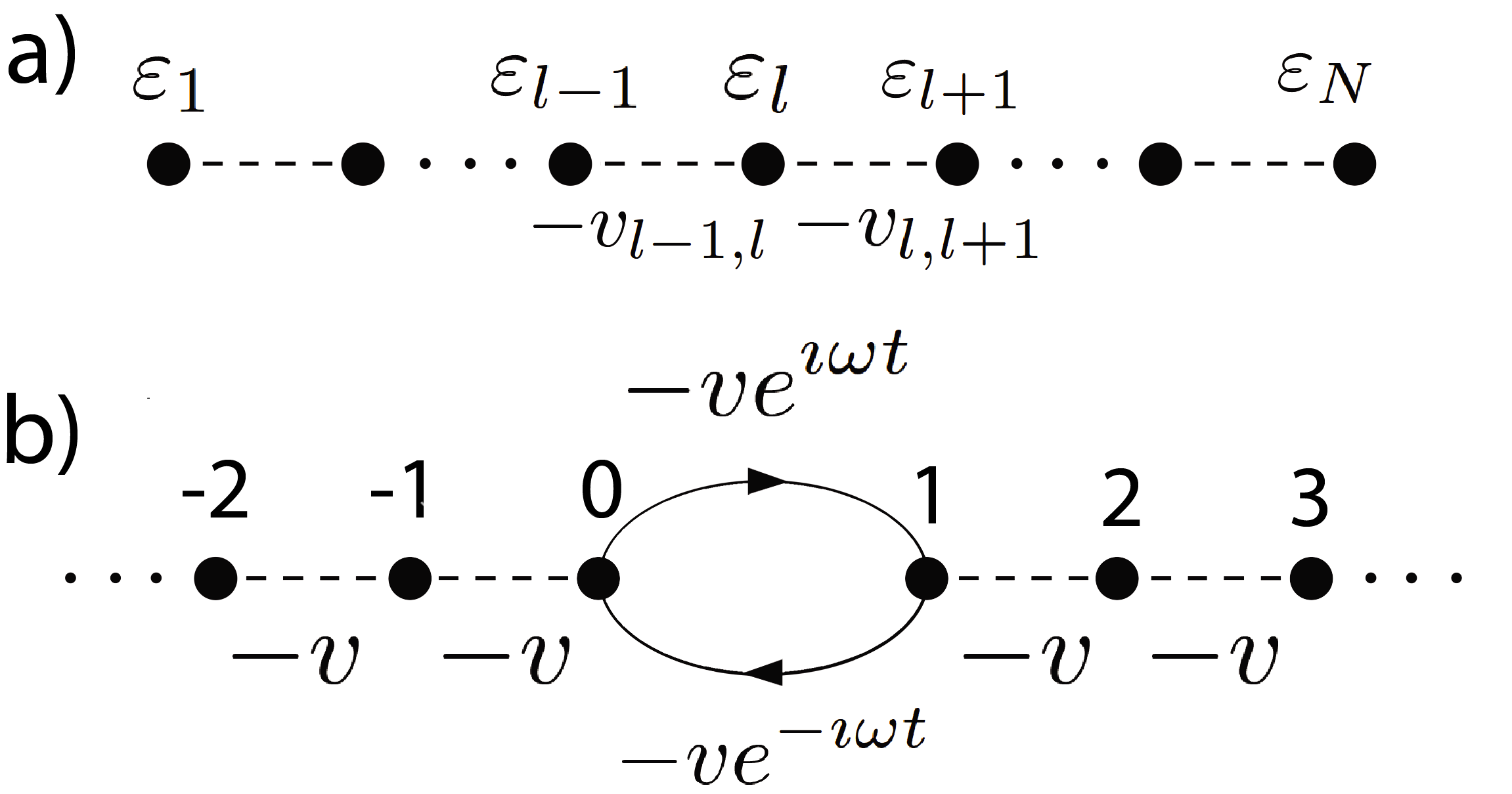}
\caption{(a) A tight-binding chain with arbitrary site energies and couplings. (b) Scattering on a time-dependent region,
i.e. on the driven coupling between sites 0 and 1, with site
energies $\epsilon_0, \epsilon_1$ }
\label{fig0}
\end{center}
\end{figure}
%---------------------------------------------------------------------------------------------------------------------------------

This observation enables us to propose a driving protocol which creates an arbitrary effective (static) potential. To this end, we consider
a tight binding chain with arbitrary site energies and couplings, see Fig. \ref{fig0}a. The sites are labeled by an integer $l$ and the site
energies are $\varepsilon_{l}$ ($l=1,\cdots,N$). The coupling constant between site $l$ and $l+1$ is $v_{l,l+1}$.

Now let us introduce driving in the couplings, in such a way that $v_{l,l+1}$ acquires a time-dependent phase $\exp\left(-\imath
\Omega_{l,l+1}t\right)$ where $\Omega_{l,l+1}$ is arbitrary. It is convenient to write $\Omega_{l,l+1}$ as a difference between two 
frequencies $i.e.,$ $\Omega_{l,l+1}=\omega_{l+1}-\omega_{l}$ (one of the frequencies, say, $\omega_{1}$ can be chosen arbitrarily
but the rest of the sequence $\omega_{2}$, $\omega_{3}$, etc, is prescribed by the values of $\Omega_{l,l+1}$). The dynamics of the
driven chain is governed by the coupled equations
\begin{align}
\imath\dot{\psi_{l}}= & \varepsilon_{l}\psi_{l}-v_{l-1,\,l}e^{\imath\left(\omega_{l}-\omega_{l-1}\right)t}\psi_{l-1}-v_{l,\,l+1}
e^{-\imath\left(\omega_{l+1}-\omega_{l}\right)t}\psi_{l+1},\label{eq: TD_eq}
\end{align}
where the dot indicates derivative with respect to time. $l$ runs from $1$ to $N$, and, e.g., the Dirichlet boundary conditions, $\psi_0
=\psi_{N+1}=0$ are imposed.

The time-dependent gauge transformation
\begin{align}
\psi_{l}\left(t\right)= & \chi_{l}\left(t\right)e^{\imath\omega_{l}t}
\end{align}
reduces Eq.~(\ref{eq: TD_eq}) to
\begin{align}
\imath\dot{\chi_{l}}= & \left(\varepsilon_{l}+\omega_{l}\right)\chi_{l}-v_{l-1,l}\chi_{l-1}-v_{l,l+1}\chi_{l+1},\label{eq: TID_eq}
\end{align}
$i.e.,$ with the help of driving, the initial Hamiltonian, defined by the sequence of site energies $\left\{ \varepsilon_{l}\right\}$ and couplings 
$\left\{ v_{l,l+1}\right\} $, is transformed to an effective static Hamiltonian, defined in Eq.~(\ref{eq: TID_eq}). This final Hamiltonian has the 
same couplings as the original one but the site energies are changed from $\left\{\varepsilon_{l}\right\} $ to $\left\{\varepsilon_{l}+\omega_{l}
\right\} $. Since the driving frequencies are at our disposal, it follows that by an appropriate choice of these frequencies it is possible to 
create any on-site potential in the equivalent static Hamiltonian. For instance, by choosing $\omega_{l}=-\varepsilon_{l}$, one can ``undo'' 
the potential $\left\{ \varepsilon_{l}\right\} $ that existed prior to driving. Or, if initially there was no potential (all $\varepsilon_{l}=0$), one 
can create by driving an arbitrary sequence of site energies, e.g., a disordered sequence which will localize the previously free excitation. 
Let us stress that {\it our driving scheme is, in general, not of a Floquet type} because the driving frequencies $\Omega_{l,\,l+1}$ are 
arbitrary and can be incommensurate with one another.

%---------------------------------------------------------------------------------------------------------------------------------
\begin{figure}
 \begin{center}
\includegraphics[width=1\columnwidth,keepaspectratio,clip]{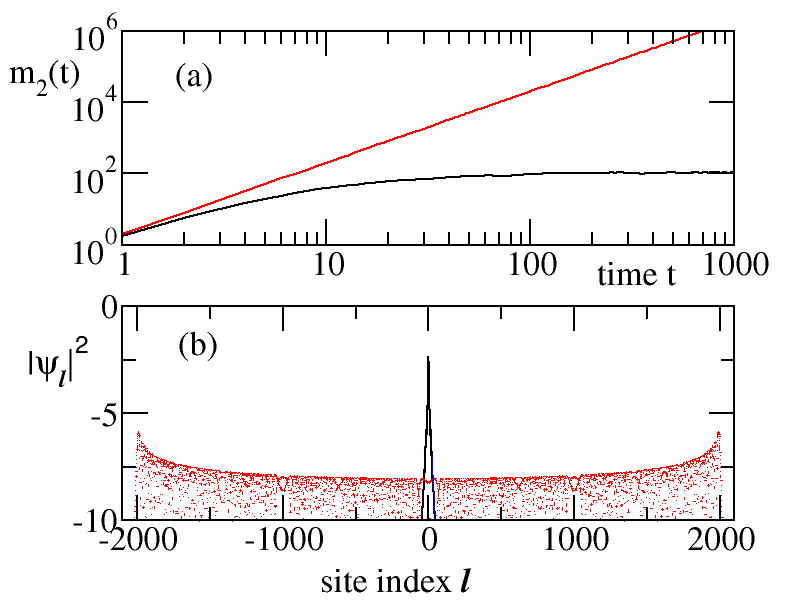}
\caption{\tp{(color online)}(a) The temporal evolution of the second moment $m_2(t)$ of a wavepacket initially localized at site $l_0=0$. 
\tp{The time is measured in units of coupling constant}. (b) The 
corresponding wavefunction profiles at the end of the simulation ($t=1000$ in units of the magnitude of the coupling constant). 
In all cases, an averaging over 100 realizations of the on-site potential and driving frequencies is performed. The black lines \tp{(lower curve in 
(a))} correspond to the case of a static on-site potential with $\epsilon_n=0$ (for all sites) and a sequence of driving frequencies
$\omega_l$ taken from a uniform distribution in the interval $\omega_l\in [0,W]$ with $W=3$. The red line \tp{(upper curve)} in (a) correspond 
to the case of $\epsilon_l\in [0,W]$ and $\omega_l=-\epsilon_l$. \tp{The wavefunction profile for the same set of parameters is indicated in
(b) with red dots.}}
\label{fig1}
\end{center}
\end{figure}
%---------------------------------------------------------------------------------------------------------------------------------

%------------------------------------
\section{Dynamics}

The metamorphosis of the wavepacket spreading in the driven tight-binding lattice can be best quantified by the 
variance of the evolving wavepacket
\begin{equation}
\label{variance}
m_2(t)\equiv\sum_l |l-l_0|^2 |\psi_l(t)|^2
\end{equation}
which at time $t=0$ is localized at site $l_0=0$, i.e. $\psi_l(t=0)=\delta_{l,0}$. The results for a lattice with (i)
random $\omega_l$ uniformly distributed in the interval $[0,W]$ and with $\epsilon_l=0$ (black line); and (ii) 
$\omega_l\in [0,W]$ and $\epsilon_l=-\omega_l$ (red line) are shown in Fig. \ref{fig1}a. In the former case we 
have induced Anderson localization via an incommensurate driving of the coupling constants. In the latter case 
we have annihilated the Anderson localization associated with a static disordered lattice by appropriate choice 
of the driving frequencies $\omega_l$. The different nature of the evolution for each of these two cases can
be further appreciated by calculating the whole evolving wavefunction intensity, see Fig. \ref{fig1}b. Again, in 
the former case (black line) we observe the familiar exponential localization of the evolving wavefunction, while 
in the latter (red dots), we find that the evolving wavefunction resembles the typical Bessel form associated
with one-dimensional translational invariant lattices (although the static potential $\epsilon_l$ was chosen to 
be random).

%---------------------------------------------------------------------------------------------------------------------------------
\begin{figure}
\begin{center}
\includegraphics[width=1\columnwidth,keepaspectratio,clip]{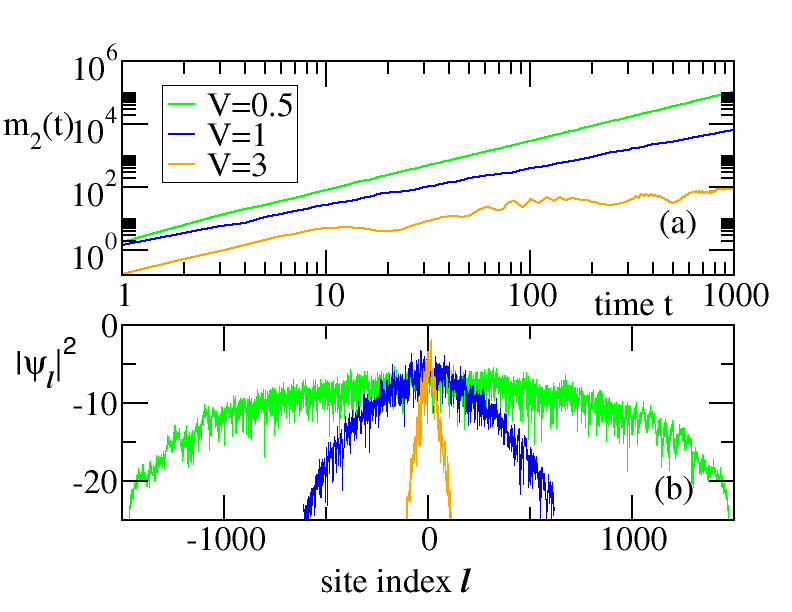}
\caption{\tp{(color online)}(a) The temporal evolution of the second moment $m_2(t)$ of a wavepacket initially localized at site $l_0=0$. Lines with different
color correspond to a set of driving frequencies $\omega_l$ taken from a Fibonacci sequence with different magnitudes $V$ \tp{($V=0.5$ upper (green) curve, 
$V=1$ middle (blue) curve and $V=3$ lower (orange) curve)}. The static on-site potential is $\epsilon_l=0$ (for all sites) in all cases. (b) The representative 
wavefunction profiles at the end of the simulation ($t=1000$ in units of the magnitude of the coupling constant).\tp{$V=0.5$ is indicated by the outer (green) 
curve, $V=1$ by the middle (blue) curve and $V=3$ by the inner (orange) curve.}}
\label{fig2}
\end{center}
\end{figure}
%---------------------------------------------------------------------------------------------------------------------------------

More exotic type of evolutions, like super-diffusion \cite{krivo}, diffusion or sub-diffusion, can be also engineered with the appropriate choice 
of the driving frequencies of the coupling constants $v_{l,l+1}$. For example, in Fig. \ref{fig2}a we show three representative cases of driving 
frequencies $\omega_l$ taken from a Fibonacci sequence with potential strengths $V=0.5$ (corresponding to super-diffusion $m_2(t) \sim 
t^{1.56}$), $V=1$ (corresponding, approximately, to diffusion $m_2(t)\sim t^{1.2}$) and $V=3$ (corresponding to sub-diffusion $m_2(t) \sim 
t^{0.6}$) \cite{guarn,geisel}. In Fig. \ref{fig2}b we show the representative wavefunction profiles at the end of the simulations ($t=1000$ in units 
of the magnitude of the coupling constant).

%-----------------------------------
\section{Scattering set-up}

We now turn to the scattering problem and start with the simple setup depicted in Fig. \ref{fig0}b 
(leaving the generalization to a scattering region of arbitrary length for later). Sites with $l\leqslant-1$ and $l\geqslant2$ represent perfect semi-infinite
leads. These sites are assigned energies (frequencies) $\varepsilon_{l}=0$ and the nearest neighbor sites are coupled by a hopping amplitude
$-v$ (we set $v=1$ to fix the energy unit). The sites $n=0,\:1$ are special. To those sites we assign energies $\varepsilon_{0},\:\varepsilon_{1}$
and couple them by the time-dependent Rabi-like coupling $-e^{\imath\omega t}$. This pair of sites constitute a scattering region from which
waves, freely propagating in the leads, get scattered. Below we show that this problem of scattering on a time-dependent scatterer has an
exact solution which provides a simple example of an {\it asymmetric frequency converter}. The field $\psi_{l}\left(t\right)$ satisfies the following
set of equations:

\begin{subequations}\label{full_eq}
\begin{align}
\imath\dot{\psi_{l}}= & -\psi_{l-1}-\psi_{l+1},\;\left(l\leqslant-1,\:l\geqslant2\right)\label{eq: dimer_scattering}\\
\imath\dot{\psi}_{0}= & -\psi_{-1}-e^{\imath\omega t}\psi_{1}+\varepsilon_{0}\psi_{0}\label{eq: eq_0}\\
\imath\dot{\psi}_{1}= & -e^{-\imath\omega t}\psi_{0}-\psi_{2}+\varepsilon_{1}\psi_{1}\label{eq: eq_1}
\end{align}
\end{subequations}

Assume that a wave $\psi_{inc}=e^{-\imath Et+\imath kl}\:\left(0<k<\pi\right)$ is incident on the scattering region from the left. $E$ and $k$
are related by the dispersion relation $E=-2\cos k$. When the wave scatters off a periodically driven scattering region, it can absorb or emit
integer number of quanta ("photons") so that the scattered wave can have frequency $E_{n}=E+n\omega$ ($n=0,\:\pm1,\cdots$). The
solution of the scattering problem posed above is of the form:

\begin{align}
\psi_{l}= & e^{-\imath Et+\imath kl}+\sum_{n}r_{n}e^{-\imath E_{n}t-\imath k_{n}l},\:l\leqslant0\label{eq: ansatz}\\
\psi_{l}= & \sum_{n}t_{n}e^{-\imath E_{n}t+\imath k_{n}l},\:l\geqslant1,\nonumber
\end{align}
where the dispersion relation $E_{n}=-2\cos k_{n}$ must hold for each $n$. If $E_{n}$ is within the band $\left(-2,\,2\right)$, the wave is
propagating, otherwise it is evanescent. Furthermore, $E_{0}=E$ and $k_{0}=k$.

The reflection and transmission amplitudes
$\left(r_{n},\:t_{n}\right)$ in Eq.~(\ref{eq: ansatz}) are
determined from the requirement that Eqs.~(\ref{eq: eq_0}) and
(\ref{eq: eq_1}) are satisfied. Using Eq.~(\ref{eq: ansatz}) and
collecting terms with the same time-dependent factors $e^{-\imath
n\omega t}$, we obtain the following set of equations:
\begin{align}
-\varepsilon_{0}-e^{\imath k}-r_{0}\left(e^{-\imath k}+\varepsilon_{0}\right)+t_{1}e^{\imath k_{1}}= & 0\label{eq: 0_1}\\
1+r_{0}-t_{1}(1+\varepsilon_{1}e^{\imath k_{1}})= & 0\nonumber
\end{align}
and
\begin{align}
\left(\varepsilon_{0}+e^{-\imath k_{n}}\right)r_{n}-t_{n+1}e^{\imath k_{n+1}}= & 0\label{eq: n_nplus1}\\
r_{n}-t_{n+1}\left(1+\varepsilon_{1}e^{\imath k_{n+1}}\right)= & 0\nonumber
\end{align}
Eqs.~(\ref{eq: 0_1}) yield
\begin{align}
r_{0}= & -\frac{1-\left(\varepsilon_{0}+e^{\imath k}\right)\left(\varepsilon_{1}+e^{-\imath k_{1}}\right)}{1-\left(\varepsilon_{0}+
e^{-\imath k}\right)\left(\varepsilon_{1}+e^{-\imath k_{1}}\right)},\label{eq: r0_t1}\\
t_{1}= & \frac{e^{\imath k}-e^{-\imath k}}{e^{\imath k_{1}}\left[1-\left(\varepsilon_{0}+e^{-\imath k}\right)\left(\varepsilon_{1}+
e^{-\imath k_{1}}\right)\right]}\nonumber
\end{align}
and from Eqs.~(\ref{eq: n_nplus1}) it follows that
\begin{align}
r_{n}=0, & \:t_{n+1}=0,\;\left(n\neq0\right).
\end{align}
Thus there is only one possibility for the incident wave $e^{-\imath Et+\imath kl}$ to get scattered into the right lead: the wave increases its
frequency from $E$ to $E+\omega$ by absorbing a photon. This process requires the condition $E+\omega=E_{1}<2$, $i.e.,$ the enhanced
frequency corresponds to the propagating wave. For $E_{1}>2$, $k_{1}$ is imaginary (evanescent wave), so that the incident wave undergoes
total reflection. Indeed, for this case, as easily verified from Eq.~(\ref{eq: r0_t1}), $\left|r_{0}\right|=1$. Since the incident frequency $E$ must
itself be in the band $\left(-2,\:2\right)$, it follows that for any $E$ the wave will be totally reflected if the driving frequency $\omega$ is larger
than $4$.

Similarly, one can consider a wave $e^{-\imath Et-\imath kl}$
impinging on the scattering region from the right. Such a wave can
be transmitted to the left with emission of a photon, so that the
emerging frequency is $E-\omega$ (provided that $E-\omega>-2$).
Thus, our scattering setup corresponds to an asymmetric frequency
converter.

It is instructive to solve the above scattering problem in a different
way, by making the transformation
\begin{align}
\psi_{l}\left(t\right)= & e^{\imath\omega t/2}\chi_{l}\left(t\right)\:\left(l\leqslant0\right);\;\psi_{l}\left(t\right)=
e^{-\imath\omega t/2}\chi_{l}\left(t\right)\:\left(l\geqslant1\right).\label{eq: scattering_transformation}
\end{align}
In the new variables, Eq.~(\ref{full_eq}) become

\begin{subequations}
\begin{align}
\imath\dot{\chi}_{l}= & \left(\frac{\omega}{2}+\varepsilon_{0}\delta_{l,0}\right)\chi_{l}-\chi_{l-1}-\chi_{l+1}\:\left(l\leqslant0\right)\\
\imath\dot{\chi}_{l}= & \left(-\frac{\omega}{2}+\varepsilon_{1}\delta_{l,1}\right)\chi_{l}-\chi_{l-1}-\chi_{l+1}\:\left(l\geqslant1\right)
\end{align}
\end{subequations}
$i.e.,$ the time-dependent driving is eliminated but the energy of the left (right) lead is raised (lowered) by $\omega/2$. The problem
is reduced to elastic scattering on a step potential (in addition to the two potential barriers at $l=0,\,1$), and it has an elementary
standard solution. In this formulation it is immediately clear that for an impinging wave $\exp\left(-\imath\tilde{E}t+\imath kl\right)$
only one wave, with the same energy parameter $\tilde{E}$, can be transmitted (reflected) to the right (left). When comparing the two
formulations, one should keep in mind that $\tilde{E}=E+\frac{\omega}{2}$, so that when going back from $\chi_{l}$ to $\psi_{l}$
(see Eq.~(\ref{eq: scattering_transformation})), one obtains $\psi_{l}\sim e^{-\imath Et}$ on the left but $\psi_{l}\sim
e^{-\imath\left(E+\omega\right)t}$ on the right, as should be.

%---------------------------------------------------------------------------------------------------------------------------------
\begin{figure}
 \begin{center}
\includegraphics[width=1\columnwidth,keepaspectratio,clip]{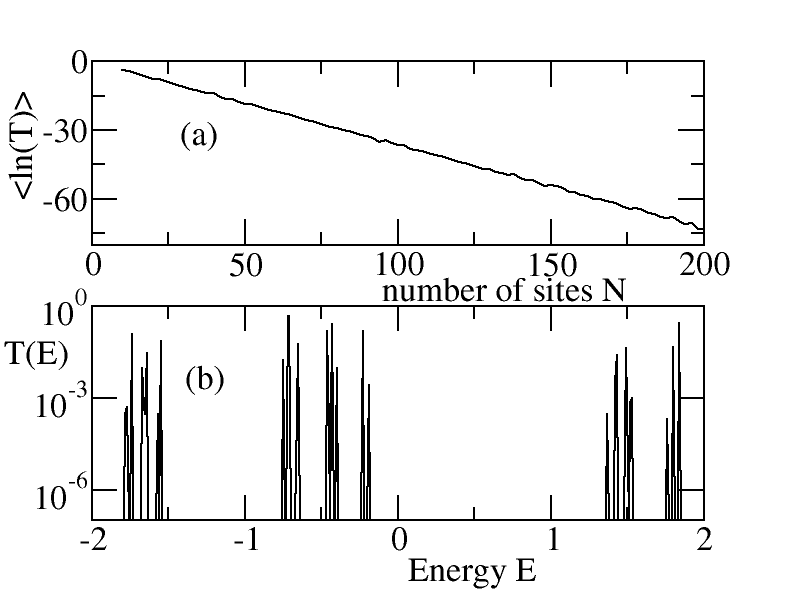}
 \caption{(a) Scaling of the logarithm of average transmittance $\langle \ln(T)\rangle$ versus the number of lattice sites $N$ for a system 
with random driving frequencies $\omega_n\in [0,W]$. In our case $W=3$. An averaging over more than 400 realizations of $\omega_n$
has been performed. (b) The transmission spectrum (single realization) versus energy $E$ \tp{(in coupling units)} of a lattice of $N=377$ 
sites with driving coupling frequencies $\omega_n$ taken from a Fibonacci sequence. The strength of the Fibonacci potential is $V=1$. 
The appearance of mini-bands is evident.}
\label{fig3}
\end{center}
\end{figure}
%---------------------------------------------------------------------------------------------------------------------------------

The advantage of the second method, based on the transformation Eq.~(\ref{eq: scattering_transformation}), is that it is easily
generalized to an arbitrary scattering sequence subjected to ``Rabi driving''. Let us return to the sequence depicted in Fig. \ref{fig0}a,
driven as explained above. This time, however, we attach perfect semi-infinite leads on both ends of the scattering sequence. Thus
the equations of motion are given in Eq.~(\ref{eq: TD_eq}), where now $l$ runs from $-\infty$ to $+\infty$ and $\omega_{l}=\omega_{1}$
for $l<1$ and $\omega_{l}=\omega_{N}$ for $l>N$ (there is no driving in the leads). It is easy to verify that the gauge transformation
\begin{align}
\psi_{l}\left(t\right)= & \chi_{l}\left(t\right)e^{\imath\omega_{l}t},\;1\leqslant l\leqslant N\\
\psi_{l}\left(t\right)=\chi_{l}\left(t\right)e^{\imath\omega_{1}t}\left(l<1\right);
&
\:\psi_{l}\left(t\right)=\chi_{l}\left(t\right)e^{\imath\omega_{N}t}\left(l>N\right)\nonumber
\end{align}
yields  the following set of equations for $\chi_{l}$:

\begin{align}
\imath\dot{\chi}_{l}= & \left(\varepsilon_{l}+\omega_{l}\right)\chi_{l}-v_{l-1,l}\chi_{l-1}-v_{l,l+1}\chi_{l+1},\:1\leqslant l\leqslant N\\
\imath\dot{\chi}_{l}= & \omega_{1\left(N\right)}\chi_{l}-\chi_{l-1}-\chi_{l+1},\:l<1\,\left(l>N\right)\nonumber
\end{align}
$i.e.,$ the scattering becomes elastic but there is an energy shift
$\left(\omega_{1}-\omega_{N}\right)$ between the two leads.

In Fig. \ref{fig3} we show some numerical simulations for two cases of driven lattices. In order to simplify our calculations we have assumed
that $\omega_1=\omega_N=0$. In this way, both leads have the same impedance (and group velocity) and thus the transmittance is simply
the square of the ratio of the transmitted to incident wave amplitude. First, we have considered a lattice with random driving frequencies
$\omega_l\in [0,W]$ and $\epsilon_l=0$. Our scaling analysis of the logarithm of the transmittance $\langle \ln(T(E))\rangle$ versus the size of
the system $N$ indicated an exponential decay -- a clear signature of Anderson localization. In Fig. \ref{fig3}b we also report the transmission
spectrum for the case of a lattice with driving frequencies $\omega_l$'s taken from a Fibonacci sequence with strength $V=1$. The formation
of mini-bands, associated with the fractal nature of the spectrum of such systems \cite{fibo,fibo2} is evident.

%-----------------------------------------------
\section{Experimental implementation} 

One might envision a number of possible experimental realizations of the time-modulated coupling scheme 
proposed above. One promising realization appears in the framework of coupled resonators which are indirectly coupled
via a set of "auxiliary" resonators. The auxiliary resonators are designed to be anti-resonant to the main resonators, i.e. their diameter is chosen 
such that the electromagnetic field is destructively interferes inside these resonators while it demonstrates constructive interferences at the main 
resonators and thus it is confined there. It turns out that in the case that the phase difference between the accumulated phases due to field 
propagation in the upper and lower sections of the auxiliary rings are not equal, the effective coupling acquires an additional phase term. 
This effect has been used originally in Ref. \cite{HDLT11,MFFMTH14} for the realization of synthetic gauge fields for photons. The same
set-up can be used in our case as well, with the addition of a phase modulator which will control these phase terms. A detail analysis of this
set-up is discussed in section A of Appendix. An alternative realization could invoke an array of resonators with a dynamically modulated nearest neighbor 
coupling $v_{l,l\pm1}\cos(\Omega_D t+\phi_{l,l\pm1})$ where $\Omega_D$ is the frequency detuning between the two resonators and $\phi_{n,n\pm1}(t)$
is a time-dependent phase of the coupling constant modulation. It turns out that in the limiting case of $v\ll \Omega_D$ (and under subsequent 
rotating wave approximations) \cite{YF15}, the system is described by Eq. (\ref{eq: TD_eq}), provided that  $\phi_{n,n\pm1}=\Omega_{n,n\pm1}t$.

%-------------------
\section{Conclusions}

In conclusion, we have proposed a new driving scheme, when different parts of a system are driven with different frequencies.
We showed that, using a particular driving protocol of this type, one can create arbitrary effective static potentials. For instance, with appropriate 
driving one can create a disordered static potential and, thus, obtain localization. The opposite is also true: a suitable driving can make any 
static potential (either deterministic or disordered) ``invisible", thus, insuring free propagation of excitations. The Hamiltonian engineering scheme
presented here can be also extended to higher dimensions (see section B of  Appendix)-- thus providing alternative pathways to realize topologically protected 
states or to induce a phase transition (like Anderson metal-insulator transition). 

\section*{Acknowledgments}

(H.L) and (T.K) acknowledge partial support by an AFOSR grant No.FA 9550-10-1-0433, and by NSF grants 
EFMA-1641109. BS acknowledges the hospitality of the Physics Department of Wesleyan University, where this work was performed. 
He is also grateful to D. Arovas, H. Herzig Sheinfux, N. Lindner and M. Segev for illuminating discussions. The authors acknowledge
an anonymous referee for valuable suggestions associated with potential realizations of the proposed driven protocol in arrays of coupled resonators.

%============================================================================================
\vspace*{2cm}
%\clearpage
%%%%%%%%%% Merge with supplemental materials %%%%%%%%%%
\pagebreak
\widetext
\begin{center}
\textbf{\large Appendix}
\end{center}

%%%%%%%%%% Merge with supplemental materials %%%%%%%%%%
%%%%%%%%%% Prefix a "S" to all equations, figures, tables and reset the counter %%%%%%%%%%
\setcounter{equation}{0}
\setcounter{figure}{0}
\setcounter{table}{0}
\setcounter{page}{1}
\makeatletter
\renewcommand{\theequation}{A\arabic{equation}}
\renewcommand{\thefigure}{A\arabic{figure}}
\renewcommand{\bibnumfmt}[1]{[A#1]}
\renewcommand{\citenumfont}[1]{A#1}
%%%%%%%%%% Prefix a "S" to all equations, figures, tables and reset the counter %%%%%%%%%%

%-----------------------------------------------------------------------------------------------------

\subsection{Photonic structures with Rabi-like couplings}

We present a design of rabi-like effective coupling between two high-Q
ring resonators using an auxiliary ring~\cite{Hafezi,Mittal,Longhi}.
On the one hand, waves in the high-Q ring resonators need to satisfy
the resonance condition. On the other hand, for our purpose the auxiliary
ring is driven.  And, at the same time as an indirect coupler, it should
satisfy the anti-resonance condition. Thus the field will be confined
mainly in the high-Q ring resonators instead of the auxiliary ring.
Below we provide a detailed analysis which serves to demonstrate this
subtle anti-resonate issue for the auxiliary ring in the case of driving.
Essentially we can generalize the derivation in Ref.~\cite{Longhi}
to the case of time-dependent couplings. 

%-----------------------------------------------------------------------------------------
\begin{figure}[h]
\includegraphics[width=0.6\columnwidth,keepaspectratio,clip]{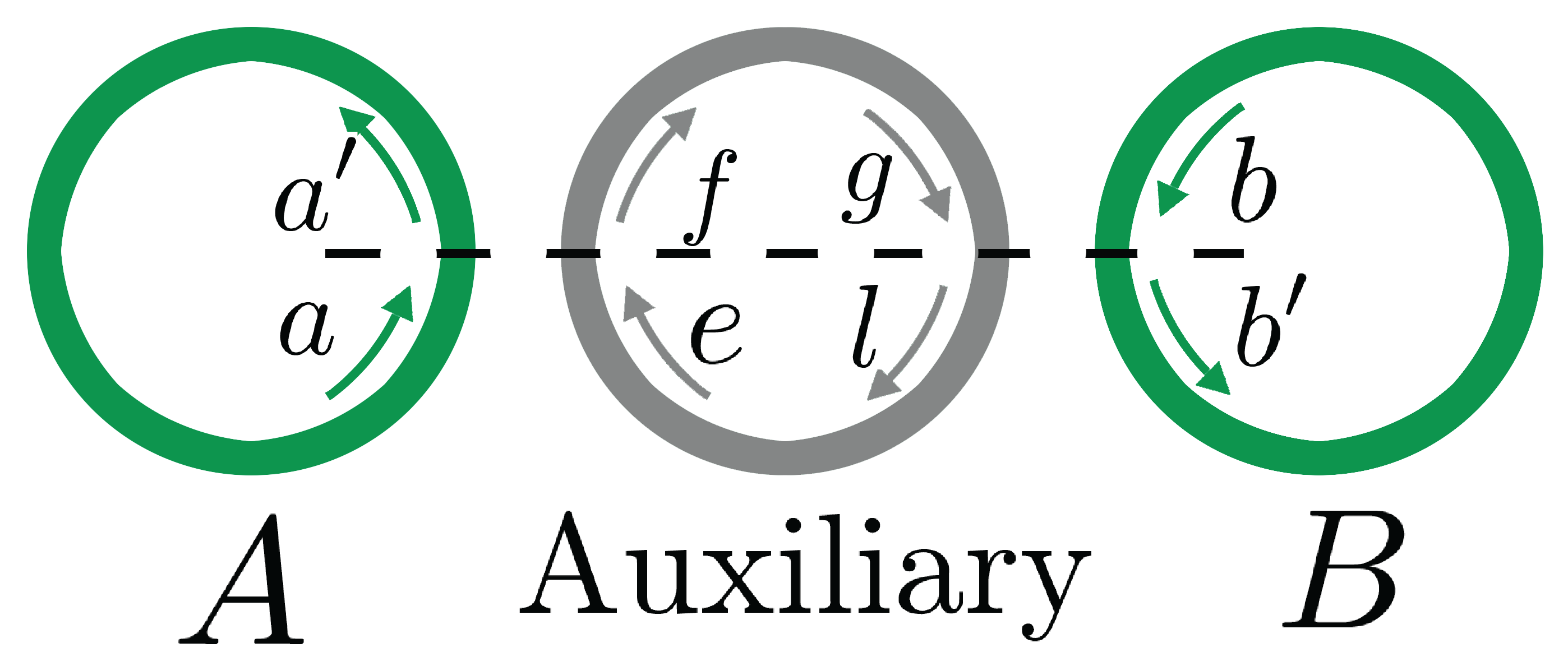}
\caption{ (color online)  The main ring resonators A and B are indirected coupled via an auxiliary ring.
The time-dependent phase modulators (not shown) embedded in both upper and lower branches of the auxiliary ring are introduced to realized effective rabi-like coupling between the resonators A and B.
 }
%\label{fig3}
\end{figure}
%----------------------------

Specifically, see Fig.~S1, we consider the field amplitudes
$a\left(t\right)$, $a'\left(t\right)$ and $b\left(t\right)$, $b'\left(t\right)$
in the two high-Q ring resonators A and B and denote the field amplitudes
$e\left(t\right)$, $f\left(t\right)$, $g\left(t\right)$, $l\left(t\right)$
in the middle auxiliary ring. The couplings between the high-Q resonators
and the auxiliary ring are given as~\cite{Yariv}
\begin{align}
\begin{pmatrix}a'\\
f
\end{pmatrix}= & \begin{pmatrix}u & \rho\\
-\rho^{*} & u^{*}
\end{pmatrix}\begin{pmatrix}a\\
e
\end{pmatrix},\begin{pmatrix}b'\\
l
\end{pmatrix}=\begin{pmatrix}u & \rho\\
-\rho^{*} & u^{*}
\end{pmatrix}\begin{pmatrix}b\\
g
\end{pmatrix}\label{eq: coupler}
\end{align}
where we assume the weak coupling limit $\left|\rho\right|\rightarrow0$
and for simplicity $u\in\mathcal{R}$. In addition, the current conservation
requires $\left|u\right|^{2}+\left|\rho\right|^{2}=1$ and thus $u\approx1$
in the weak coupling limit. The ring resonators A and B are considered
to be lossless and the resonance condition provides us that 
\begin{align}
a\left(t+\tau\right)= & a'\left(t\right),b\left(t+\tau\right)=b'\left(t\right),\label{eq: BCs}
\end{align}
where $\tau$ is the traveling time when wave goes around the ring
A or B for one circle. More importantly, for the middle lossless auxiliary
ring, we have 
\begin{align}
g\left(t+\tau_{1}\right)=f\left(t\right)\exp\left[\imath\phi_{1}\left(t\right)\right], & e\left(t+\tau_{1}\right)=l\left(t\right)\exp\left[\imath\phi_{2}\left(t\right)\right],\label{eq: Modulator}
\end{align}
where the propagation time in the half auxiliary ring is $\tau_{1}\sim\mathcal{O}\left(\tau\right)$,
the effective time-dependent phases are assumed to be $\phi_{1}\left(t\right)=\frac{\pi}{2}+\omega t$
and $\phi_{2}\left(t\right)=\frac{\pi}{2}-\omega t$, with $\omega\ll 1/\tau_1$, which are due to the phase
modulators in both upper and lower branches of the auxiliary ring.
Notice that the anti-resonance condition holds for the auxiliary ring
irrespective of the time instant, $i.e.,$ $\phi_{1}\left(t\right)+\phi_{2}\left(t\right)=\pi$.

Assuming small traveling time for one circle, $i.e.,$ $\tau\rightarrow0$,
we have 
\begin{align}
a\left(t+\tau\right)\approx & a\left(t\right)+\tau\frac{da}{dt},b\left(t+\tau\right)\approx b\left(t\right)+\tau\frac{db}{dt}.\label{eq: mean_F}
\end{align}
Using Eqs.~(\ref{eq: coupler}) and (\ref{eq: BCs}), we get from
Eq.~(\ref{eq: mean_F})

\begin{align}
\tau\frac{da}{dt}\approx-\left(1-u\right)a+\rho e, & \tau\frac{db}{dt}\approx-\left(1-u\right)b+\rho g.\label{eq: intermidiate}
\end{align}
We proceed to eliminate field amplitudes $e$ and $g$ in Eq.~(\ref{eq: intermidiate}).
To this end, combining Eqs.~(\ref{eq: coupler}) and (\ref{eq: Modulator}),
up to the leading order with respect to the small quantities $\left|\rho\right|$
and $\tau$ we get 
\begin{align}
g\approx & \frac{\rho^{*}}{2}\left[ub-a\exp\left(\imath\phi_{1}\right)\right],e\approx\frac{\rho^{*}}{2}\left[ua-b\exp\left(\imath\phi_{2}\right)\right].\label{eq: g_e}
\end{align}
Resulting from the anti-resonance condition, indeed the field amplitudes
$e$ and $g$ in the auxiliary ring is small in the weak coupling
limit, see Eq.~(\ref{eq: g_e}). Now substituting Eq.~(\ref{eq: g_e})
into Eq.~(\ref{eq: intermidiate}), finally we get
\begin{align}
\imath\frac{da}{dt}= & \frac{\kappa}{\tau}\exp\left(-\imath\omega t\right)b,\imath\frac{db}{dt}=\frac{\kappa}{\tau}\exp\left(\imath\omega t\right)a,\label{eq: rabi}
\end{align}
where $\kappa\equiv\left|\rho\right|^{2}/2$. The rabi-like coupling
arises in Eq.~(\ref{eq: rabi}).

\subsection{Generalization of the driving protocol to two-dimensions}

%-----------------------------------------------------------------------------------------
\begin{figure}[h]
\includegraphics[width=0.5\columnwidth,keepaspectratio,clip]{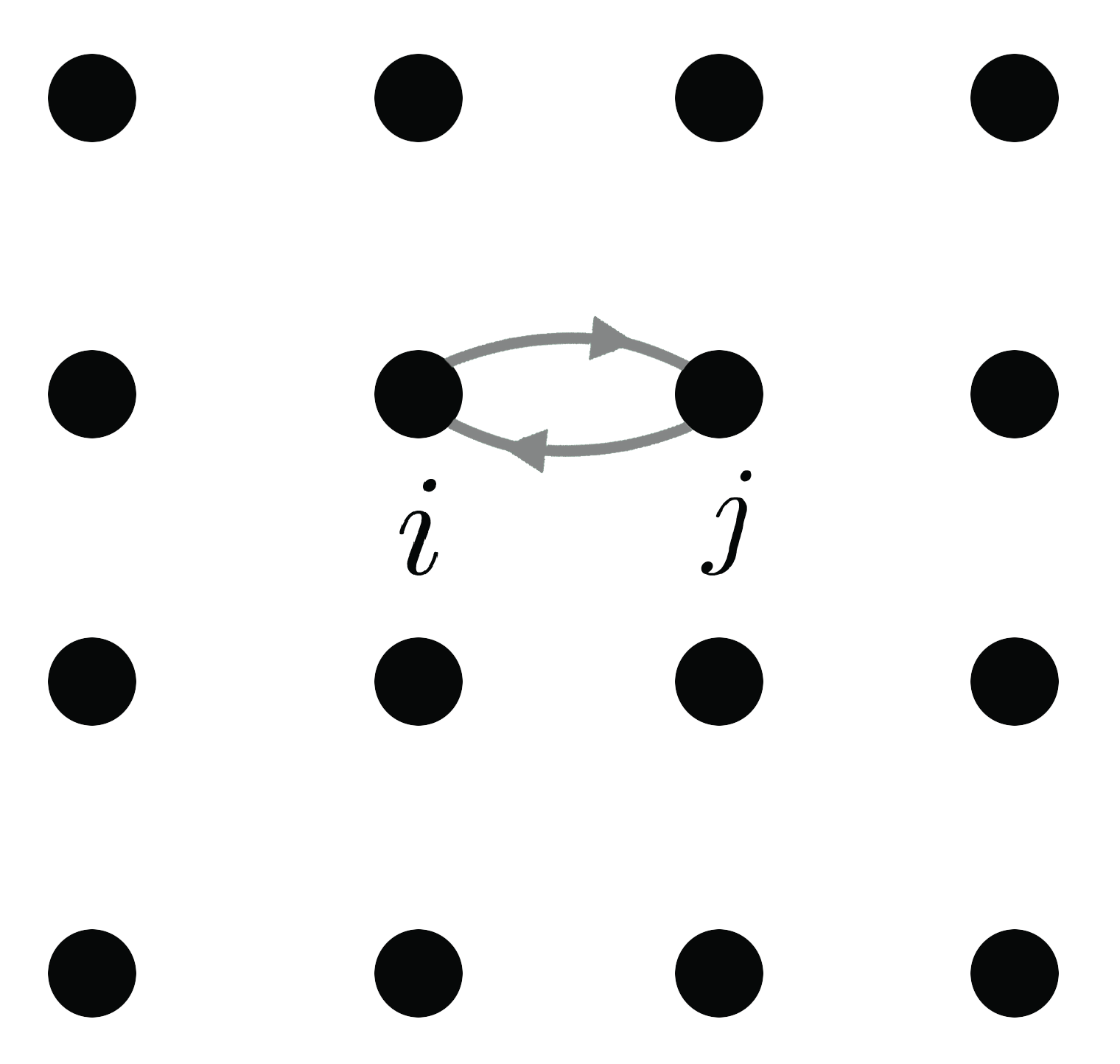}
\caption{ (color online)  Schematic of a 2D square lattice. The rabi-like effective coupling is shown explicitly for two nearest site $i$ and $j$.  }
%\label{fig3}
\end{figure}
%----------------------------

Consider any site $i$ on a  two-dimensional (2D) lattice, see Fig.~S2. The set of four nearest neighbors is labeled by $j$. The equation for $\psi_i$, before driving, is
\begin{equation}
\label{tbebefore}
\imath{\dot \psi}_i =\epsilon_i \psi_i -V\sum_{j\in n.n.} \psi_j
\end{equation}
where the sum is over nearest neighbors. More generally one could have $V_{i,j}$ instead of $V$.

Now we introduce driving according to the following rule: each site is assigned a frequency $\omega_i$ and the bond $(i,j)$ is driven as $-V e^{\imath(\omega_i
-\omega_j)t}$. thus we have:
\begin{equation}
\label{tbeafter}
\imath{\dot \psi}_i =\epsilon_i \psi_i -V\sum_{j\in n.n.} e^{\imath(\omega_i-\omega_j)t} \psi_j 
\end{equation}
One can eliminate driving by the gauge transformation Eq. (4) (see main text) and come up with the following equation:
\begin{equation}
\label{tbeafter}
\imath{\dot \chi}_i =(\epsilon_i+\omega_i) \chi_i -V\sum_{j\in n.n.}  \chi_j 
\end{equation}
which is a generalization of Eq. (5) (see main text) in 2D. A further generalization to three-dimensions is obvious and it is not discussed here. 
It is, nevertheless, important to stress that the Rabi-type of driving that we discuss here can only create an effective static scalar potential $\omega_i$ at 
site $i$ and cannot ``generate" any synthetic magnetic field. Indeed by denoting the phase $(\omega_i-\omega_j)t\equiv \phi_{i,j}(t)$, we can show that the 
sum of $\phi_{i,j}(t)$ on any close contour of the lattice is zero, at all times.

\end{document}